\documentclass[aps,prl,preprint,superscriptaddress]{revtex4}

\usepackage{graphicx}
\usepackage{amssymb}

\begin{document}

\title{High-Multipolar Effects on Dispersive Forces} 

\author{Cecilia Noguez}
\email[Corresponding author. Email:]{cecilia@fisica.unam.mx}
\affiliation{Instituto de F\'{\i}sica, Universidad Nacional Aut\'onoma
 de M\'exico, Apartado Postal 20-364, D.F. 01000,  M\'exico}

\author{C. E. Rom\'an-Vel\'azquez}
\affiliation{Centro de Investigaci\'on en Ciencia Aplicada y Tecnolog\'{\i}a 
Avanzada, Instituto Polit\'ecnico Nacional, Av. Legaria 694, Col. 
Irrigaci\'on, D.F. 11500, M\'exico}

\author{R. Esquivel-Sirvent}
\affiliation{Instituto de F\'{\i}sica, Universidad Nacional Aut\'onoma
 de M\'exico, Apartado Postal 20-364, D.F. 01000,  M\'exico}

\author{C. Villarreal}
\affiliation{Instituto de F\'{\i}sica, Universidad Nacional Aut\'onoma
 de M\'exico, Apartado Postal 20-364, D.F. 01000,  M\'exico}

\date{\today}

\begin{abstract}
We show that the dispersive force between a spherical nanoparticle (with a radius $\le$ 100~nm) and a substrate is enhanced by several orders of magnitude when the sphere is near to the substrate. We calculate exactly the dispersive force in the non-retarded limit by incorporating the contributions to the interaction from of all the multipolar electromagnetic modes. We show that as the sphere approaches the substrate, the fluctuations of the electromagnetic field, induced by the vacuum and the presence of the substrate, the dispersive force is enhanced by orders of magnitude. We discuss this effect as a function of the size of the sphere.     
 \end{abstract}


\maketitle

The technical and experimental advances associated to the investigation of micrometric and nanometric devices has boosted the research on dispersive forces, like Casimir and van der Waals forces~\cite{ninham}, which arise from  fluctuations of the quantum vacuum due to boundaries or external fields. As it is well known, Casimir~\cite{casimir} predicted an attractive force between two perfect conductors parallel plates separated by a given distance. The Casimir theory was extended latter to dielectric materials by Lifshitz in 1956~\cite{lifshitz}. Due to the experimental difficulties inherent in handling two parallel plates at small distances, Derjaguin and collaborators measured the Casimir force between a glass plate and a relatively large spherical lens~\cite{proximidad} and they developed the so called Dejarguin approximation or proximity theorem. This theorem allows the estimation of the Casimir force per unit area between a sphere and a plate if the energy per unit area between two parallel plates is known. However, this theorem only holds when the radius $R$ of the sphere and the distance $z$ between it and the substrate satisfies the relation, $z/R \ll 1$. Most of the accurate measurements of the Casimir forces performed in recent times rely upon this theorem~\cite{lamoraux,chan,mohideen}. However, it is not clear up to what limit this approach remains valid, and of course, for $R \sim z$, an exact calculation of the Casimir force between a sphere and a planar surface becomes essential.

Within the Casimir effect context, the problem of a particle over a substrate reduces to finding the effects of the environment acting on the particle. This is not a simple problem since it must be solve in a self-consistent fashion, since the environment, in this case the substrate, has an effect on the particle and the particle also affects the environment. 
In a previous work~\cite{ceci}, the optical properties of a spherical particle lying above a substrate have been determined through its response to a local electromagnetic field. In this case, this field is produced by an external or applied field, plus a field coming from the presence of the substrate. The latter was introduced in the problem as a charge distribution induced on the particle due to the presence of the substrate which modifies the response of the particle itself. 
This procedure is implemented by considering the interaction of the induced dipole moment of the particle with its own image dipole in the substrate. However, this approach fails to study the optical properties of nanoparticles that are very close to the substrate, such that the ``image-dipole'' model must be extended in order to include higher-order multipolar moments. Previous extensions of the ``image-dipole'' model have shown that the importance of these multipolar excitations increases with the proximity of the particle to the substrate~\cite{ceci}. This ``image-dipole'' model was also considered by Casimir and Polder, in 1948, to find the London or van der Waals forces of a polarizable atom near a perfect conductor plane~\cite{casypol}. They found a correction to the van der Waals force by considering the influence of retardation. In 1998, Ford~\cite{ford} also considered the ``image-dipole'' model to calculate the Casimir Force between a perfectly conducting wall and a large sphere with a Drude dielectric function. In early experiments to measure the Casimir-Polder interaction, Shih and Parsegian~\cite{shih} measured the deflection of atomic beams by gold surfaces obtaining results that are inconsistent with the ``image-dipole'' model. The measurements of Shih and Parsegian~\cite{shih}, as well as other experimental observations~\cite{lafyatis}, indicate that the Casimir or van der Waal force between a polarizable particle and a planar substrate should involve more complicated interactions than the simple ``image-dipole'' model. 

In this paper, we calculate exactly the dispersive force, or Casimir force in the non-retarded limit, between a  nanosphere and a plane. The force is obtained considering all the high-multipolar excitations in system, that are shown to enhance the force by several orders of magnitude as a nanosphere approaches the substrate. We use a spectral representation (SR) method~\cite{ceci2} to calculate the force, that yields the frequencies ($\omega_s$) of the proper electromagnetic modes of the sphere-substrate system trough a diagonalization process. Therefore, the zero-point energy is easily calculated as ${\cal E} = \sum_s 1/2 \hbar \omega_s$. Another advantage of the SR is that it separates the contribution of the dielectric properties of the sphere from the contribution of its geometrical properties. Both properties of the SR allows to perform systematic studies of dispersive forces in this system. 

We consider a nanometric-size sphere of radius $R$ located at a a minimum distance $z$($\le R$) from a substrate. In this case, the radius of the sphere and the minimum separation between the sphere and the substrate, are smaller than the characteristic length of the system~\cite{longitud}, therefore, it is not necessary to consider retardation effects.
The fluctuations of the electromagnetic field induce a multipolar moment which also induces a charge distribution in the substrate that affects the multipolar moment on the sphere, trough a self-induced polarization process. Then, the $lm$-th multipolar moment on the sphere, $Q_{lm}$, is induced by a local electromagnetic field, $V_{lm}^{\rm loc}$, which is the sum of the induced field due to the quantum vacuum fluctuations at the zero-point energy ($V_{lm}^{\rm vac}$), plus the induced field due to the presence of the substrate ($V_{lm}^{\rm sub})$. Therefore, the $Q_{lm}$ of a sphere with polarizabilities $\alpha_{lm}(\omega)$ is given by~\cite{claro}
 \begin{eqnarray}
Q_{lm}  &=& - \frac{(2l +1)}{4 \pi} \alpha_{lm} (\omega) V_{lm}^{\rm loc}, \\
 &=& -\frac{(2l+1)}{4 \pi} \alpha_{lm}(\omega) \left[ V_{lm}^{\rm vac} + V_{lm}^{\rm sub} \right], \nonumber 
\end{eqnarray}
where the $lm$-th multipole moment induced in the sphere is defined as
\begin{equation}
Q_{lm}=\int_{v_s} r'^l \rho({\bf r'}) Y_{lm}(\theta', \varphi')\,d^3r',  \label{Qlm}
\end{equation}
where $\rho ({\bf r'})$ is the charge density in the sphere, $Y_{lm}$ are the spherical harmonics, and the integral is performed over the volume $v_s$ of the sphere. The $lm$-th component of the local field can be written using a multipolar expansion~\cite{claro} that yields to
\begin{equation}
Q_{lm} = \frac{-(2l+1)}{4 \pi} \alpha_{lm}(\omega) \left[ V_{lm}^{\rm vac} + \sum_{l', m'} (-1)^{m^{^{\prime }}+l^{^{\prime }}}  A_{lm}^{l'm'} {\hat{Q}}_{l'm'} \right],  \label{q}
\end{equation}
where $\hat{Q}_{l'm'}$ is the $l'm'$-th induced multipolar moment in the substrate  which is located at ${\bf r}= (2(z+R),\theta=\pi,\varphi)$ from the center of the sphere, and $A_{lm}^{l'm'}$ is the matrix that couples the interaction between the multipolar distribution on the sphere and substrate~\cite{claro}. The induced $l'm'$-th multipolar moment in the substrate or the ``image-multipole'' moment  is related with the $Q_{lm}$ on the sphere by 
\begin{equation}
\hat{Q}_{l'm'} = (-1)^{l'+m'} f_c(\omega) Q_{l'm'},
\end{equation}
where $f_c(\omega)$ is a contrast function that together with the factor $(-1)^{l'+m'}$ allows to satisfy the boundary conditions of the electromagnetic field on the sphere-substrate system~\cite{fc}. Here, we have assumed that the substrate is electrically neutral and has a local dielectric function $\epsilon_{\rm sub}(\omega)$, and the ambient between the sphere and substrate is the vacuum. Let us suppose that the sphere is homogeneous, and has a local dielectric function $\epsilon_{\rm sph} (\omega)$, such that it has azimuthal symmetry, and its polarizabilities are independent of $m$, and are given by~\cite{claro} 
\begin{equation}
\alpha_l (\omega) = \frac{l[\epsilon_{\rm sph}(\omega) - 1]} {l [ \epsilon_{\rm sph}(\omega) + 1] + 1} R^{2l+1}. \label{alfa}
\end{equation}
To find the proper electromagnetic modes of the sphere-substrate system, we need to solve Eq.~(3) by a self-consistent procedure or by inverting $L(2L+1)$ complex matrices of $L\times L$ dimension, where $L$ is the largest order of the multipolar expansion. To avoid these cumbersome procedures, we use instead a spectral representation that allows to find the proper electromagnetic modes trough the diagonalization of a real and symmetric matrix. The discussion of the above method can be found in Ref.~13.

First, we rewrite the sphere polarizabilities from Eq.~(\ref{alfa}) as
\begin{equation}
\alpha_l (\omega) =  \frac{n_0^l}{n_0^l - u(\omega)} R^{2l+1}, \label{alfa2}
\end{equation}
where $u(\omega) = [1 - \epsilon_{\rm sph}(\omega)]^{-1}$ and $n_0^l = l/(2l+1)$. One should notice that the poles of  Eq.~(\ref{alfa2}), that is $u(\omega_l) = n_0^l$, yield the frequencies of the proper electromagnetic modes of the isolated sphere, therfore $u(\omega)$ is known as a spectral variable~\cite{ceci}. Now,  we rewrite Eq.~(\ref{q}) as an eigenvalue equation as
\begin{equation} 
\sum_{\mu'} \left[ - u(\omega) \delta_{\mu \mu'}  +  {H}^{\mu '}_{\mu}  \right] {x}_{\mu'} = g_{\mu}, \label{h2}
\end{equation}
where we have simplified the notation by writing $\mu \equiv (l,m)$, and 
\begin{equation}
x_{\mu} = \frac{Q_{lm}} { (l R^{2l+1})^{1/2}}, \quad \quad 
g_{\mu} = - \frac{(l R^{2l+1})^{1/2}} {4 \pi} V_{lm}^{\rm vac},
\end{equation}
and
\begin{equation}
H_{\mu}^{\mu'} = n_{l'}^0  \delta_{\mu \mu'} + f_c(\omega) \frac{(R^{l+l'+1})} {4 \pi} (ll')^{1/2} A_{\mu}^{\mu'}. \label{h}
\end{equation}
It is shown in Ref.~15 that $A_{\mu}^{\mu'}$ is a symmetric matrix that depends on the distance between the ``image-multipole'' and the sphere as, $1/[2(z+R)]^{l+l'+1}$, such that $H_{\mu}^{\mu'}$ is dimensionless and symmetric, and only depends on the geometry of the system through the ratio $z/R$. Consider the case when the contrast factor function $f_c(\omega)$ is real~\cite{rota}, then $H_{\mu}^{\mu'}$ is also real, and we can always find a unitary transformation that diagonalizes it, such that  $ \sum_{\mu \mu'}  (U_{\nu}^{\mu})^{-1}H_{\mu}^{\mu'}U_{\mu'}^{\nu'}  =  4 \pi n_{\nu} \delta_{\nu \nu'}$, being $n_{\nu}$ the eigenvalues of the matrix $H_{\mu}^{\mu'}$. Then, the solution of Eq.~(\ref{h2}) is given by
\begin{equation}
x_{\mu} = - \sum_{\mu'} G_{\mu}^{\mu'} g_{\mu'},
\end{equation}
where $ G_{\mu}^{\mu'} $
is a Green's operator, whose $\mu \mu'$ element can be written in terms of the $U_{\nu}^{\mu}$ and the eigenvalues of Eq.~(\ref{h}), as
\begin{equation}
G^{\mu}_{\mu'}(u) = G^{ml}_{m'l'} = \sum_{m''l''}  \frac{U_{m''l''}^{ml}(U_{m'l'}^{m''l''})^{-1}}{u-n_{m''}^{l''}}. \label{green}
\end{equation}
The poles of $G^{ml}_{m'l'}(u)$, that is $u(\omega^{l}_{m}) = n_{m}^{l}$, yield the frequencies of the proper electromagnetic modes of the system. If the sphere is isolated there is no substrate, therefore, the dielectric function of the substrate is replaced by the one of the vacuum and $f_c(\omega) = 0$, such that  $n_{m}^{l} = n_0^l$.

We now calculate the interaction energy as the difference ${\cal E}(z/R) = {\cal E}_{z/R} - {\cal E}_\infty $, where ${\cal E}_{z/R} $ is the energy when the sphere of radii $R$ is at a distance $z$ from the substrate, and ${\cal E}_\infty$ is the energy when the sphere is far from the substrate ($z \to \infty$) or isolated, that is,
\begin{equation}
{\cal E}(z/R) = \sum_{lm} \frac{\hbar \omega_{m}^l}{2} - \sum_{l} \frac{\hbar \omega_0^{l}}{2}. \label{ener}
\end{equation}

The results presented here are calculated as follows. First, we construct the matrix $H$ for a given $z/R$, choosing a maximum value of multipolar excitations that ensures convergence in the eigenvalues and eigenvectors of $H$, and then, we diagonalize it. We consider an explicit expression for the dielectric function of the sphere, and then we calculate the proper electromagnetic modes  trough the relation $u(\omega^{l}_{m}) = n_{m}^{l}$. Once we have $\omega^l_m$, we calculate the energy according with Eq.~(\ref{ener}). The largest order of the multipolar expansion considered in this work was $L=2000$ at the minimum distance of $z/R = 0.001$, because multipolar modes of higher order than $L$ do not contribute to the energy ${\cal E}(z/R)$ at such distance. Here, we use the Drude model for the dielectric function of the sphere, therefore, $u(\omega) =  [\omega(\omega + i/\tau)]/\omega_p^2$, where $\omega_p$ is the plasma frequency and $\tau$ is the relaxation time. We present results for aluminum (Al) spheres  with $\hbar \omega_p = 15.80$~eV, and $(\tau \omega_p)^{-1} = 0.04$. The substrate is sapphire whose dielectric function is real and constant in a wide range of the electromagnetic spectrum, with $\epsilon_{\rm sub} = 3.13$ and contrast factor $f_c(\omega)=-0.516$. 

In Fig.~1, we show the energy calculated according with Eq.~(\ref{ener}) as a function of $z/R$. We show results when all multipolar effects are taken into account (solid line), as well as results when only  quadrupolar effects (dashed line), and dipolar effects (doted line) are considered. In general, we observe that the energy shows a power law of $(z/R)^{-3}$ when only dipolar effects are considered while for the curve corresponding to quadrupolar multipoles  shows three different regions: (i) when $z/R > 7$ the energy shows a power law of $(z/R)^{-3}$, that means that only dipole-dipole interactions are important, (ii) when $ 2 < z/R < 7$ the energy shows a power law of $(z/R)^{-4}$, indicating that dipole-quadrupole interactions are important, and (iii) when $z/R < 2$ the energy shows a power law of $(z/R)^{-5}$, such that only quadrupole-quadrupole interactions become important. 


At small distances, the energy of the system can increase up to three orders of magnitude when all the contributions from high-multipolar moments are considered, as compared with the energy when only dipolar or quadrupolar moments are taken into account. Notice that when we are calculating the energy with all the high-multipolar contributions, we are solving the problem exactly. Then, the exact energy curve shows a power law of $(z/R)^{-3}$ when $z/R > 7$, when $ 2 < z/R < 7$ the energy shows the same behavior as the curve where quadrupolar effects are included, and at smaller distances ($z/R < 1$) the energy increases sharply as the minimum distance also does, and it is not possible to assign a power law. This means that high-multipolar contributions become very important as the sphere approaches the substrate. In other words, the induced field in the substrate becomes very inhomogenous when the sphere is almost touching the substrate and  then the attractive force between the sphere and the substrate increases tremendously. 


In Fig.~2, we show the Casimir force for an Al sphere over sapphire calculted as $F = - d{\cal E}/dz. $ We show the force when all multipolar interactions are taken into account, as well as up to dipolar, and up to quadrupolar interactions are considered. In agreement with the results for the energy, we observed for the force that multipolar effects become evident when the minimum distance between the sphere and the substrate is smaller than $R$. As the sphere approaches the substrate, the attractive force suddenly increases up to four orders of magnitude as compare with the dipolar interactions. At large distances ($z > 2R$) the force can be obtained exactly if only up to quadrupolar interactions are considered. We also obtained that for $z > 7R$ the interaction between the sphere and the substrate can be modeled using the dipolar approximation~\cite{ceci2}. In Fig.~3, we show the Casimir force for spheres of different radii, where all the high-multipolar modes were taken into account. We observed that when the sphere is almost touching the substrate, the attractive force is larger for the sphere with smaller radii, however, the force decays faster at larger distances when the sphere becomes smaller. Then, we observe that at a minimum distance of $z=50$~nm, the force for the sphere with $R=100$~nm is ten times larger that the force for the sphere with $R=10$~nm. Notice that at such distance only dipolar interactions are important if $R=10$~nm, while multipolar contributions are of importance if $R=100$~nm.
 
In conclusion, we have shown that the dispersive or Casimir force is enhanced by several orders of magnitude as a nanoparticle approaches a substrate. We have calculated exactly the interaction energy between the nanoparticle and the substrate using a spectral representation formalism. We found that as the separation between the particle and the substrate decreases, more multipolar moments contribute to the energy. On the other hand, for large separations the dipolar term is the dominant one, like in the Casimir and Polder model. The radius of the nanoparticle plays an important role, since at small distances the force increases if the radius decreases, while at large distances the contrary occurs. The increment of the force at small separations could explain the physical origin of the large deviations observed in the deflection of atomic beams by metallic surfaces. However, specific experiments have to be perform to prove the this.

\begin{acknowledgments}
This work has been partly financed by CONACyT grant No.~36651-E and by DGAPA-UNAM grants No.~IN104201 and IN107500.
\end{acknowledgments}

 \begin{figure}[tbh]
\centerline {
\includegraphics{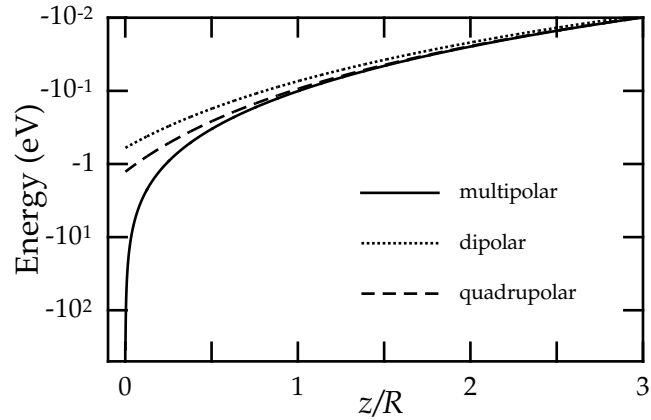}
} \vskip-.2cm
\caption{Casimir energy as a function of the ratio $z/R$.}
 \end{figure}

 \begin{figure}[tbh]
\centerline {
\includegraphics{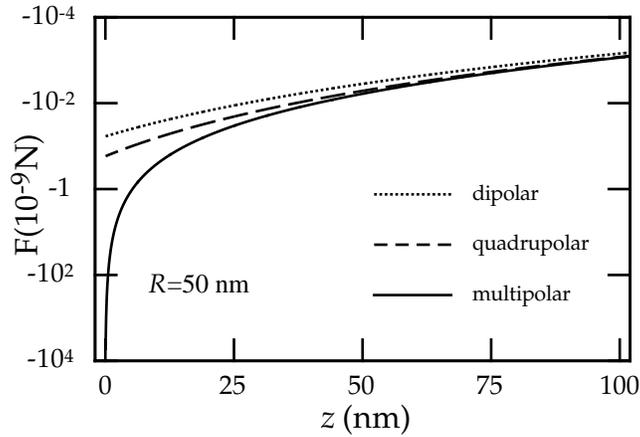}
}
\caption{Casimir force as a function of the minimum distance $z$ for a sphere with $R=50$~nm.}
 \end{figure}

\begin{figure}[tbh]
\centerline {
\includegraphics{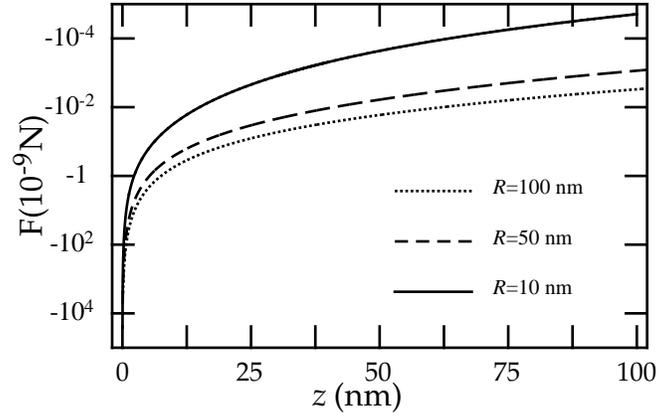}
} \vskip-.2cm
\caption{Casimir force as a function of the minimum distance $z$ for spheres with $R=10$, $R=50$, and $R=100$~nm.}
 \end{figure}


\begin{references}
\bibitem{longitud} The characteristic length of the system is given, in this case, by $l=c/\omega_p$, with $c$ the speed of light, and $\omega_p$  the plasma frequency of the sphere or the substrate.
\bibitem{rota} If the complex dielectric function of the substrate is described by the Lorentz or Drude formula, we can perform a rotation to the imaginary axis  ($\omega \to i\omega$) to get a real dielectric function.
\end{references}
\end{document}